\documentclass[final,5p,times,twocolumn]{elsarticle}

\usepackage{color}
\usepackage{amsmath,amssymb}

\newcommand{\bra}[1]{\ensuremath{\left\langle#1\right|}}
\newcommand{\ket}[1]{\ensuremath{\left|#1\right\rangle}}
\newcommand{\fig}[1]{figure~\ref{fig:#1}}
\newcommand{\Fig}[1]{Figure~\ref{fig:#1}}
\newcommand{\tab}[1]{table~\ref{tab:#1}}

\renewcommand{\sec}[1]{section~\ref{sec:#1}}

\newcommand{\eq}[1]{(\ref{eq:#1})}

\newcommand{\Abs}[1]{\ensuremath{\left| #1 \right|}}
\newcommand{\Tr}[1]{\ensuremath{\mathrm{Tr}\left(#1\right)}}

\renewcommand{\Im}[1]{\ensuremath{\mathrm{Im} \left(#1\right)}}
\newcommand{\I}{\mathrm{i}}
\newcommand{\Sg}{\Sigma}
\newcommand{\sg}{\sigma}
\newcommand{\eps}{\epsilon}
\newcommand{\angstrom}{\textup{\AA}}
\newcommand{\un}[1]{\ensuremath{\,\mathrm{#1}}}

\begin{document}

\begin{frontmatter}

\title{Microwave emulations and tight-binding calculations of transport in polyacetylene}

\author[1]{Thomas Stegmann}
\author[2,1]{John A. Franco-Villafa\~ne}
\author[1]{Yenni P. Ortiz}
\author[3]{Ulrich Kuhl}
\author[3]{Fabrice Mortessagne}
\author[1,4]{Thomas H. Seligman}

\address[1]{Instituto de Ciencias F\'isicas, Universidad Nacional Aut\'onoma de M\'exico,
  Avenida Universidad s/n, 62210 Cuernavaca, M\'exico}
\address[2]{Instituto de F\'isica, Benem\'erita Universidad Aut\'onoma de Puebla,
  Apartado Postal J-48, 72570 Puebla, M\'exico}
\address[3]{Universit\'e de Nice - Sophia Antipolis, Laboratoire de la Physique de la Mati\`ere
  Condens\'ee, CNRS, Parc Valrose, 06108 Nice, France}
\address[4]{Centro Internacional de Ciencias, 62210 Cuernavaca, M\'exico}

\begin{abstract}
  A novel approach to investigate the electron transport of \textit{cis}- and
  \textit{trans}-polyacetylene chains in the single-electron approximation is presented by using
  microwave emulation measurements and tight-binding calculations. In the emulation we take into
  account the different electronic couplings due to the double bonds leading to coupled dimer
  chains. The relative coupling constants are adjusted by DFT calculations. For sufficiently long
  chains a transport band gap is observed if the double bonds are present, whereas for identical
  couplings no band gap opens. The band gap can be observed also in relatively short chains, if
  additional edge atoms are absent, which cause strong resonance peaks within the band gap. The
  experimental results are in agreement with our tight-binding calculations using the nonequilibrium
  Green's function method. The tight-binding calculations show that it is crucial to include third
  nearest neighbor couplings to obtain the gap in the cis-polyacetylene.
\end{abstract}


\end{frontmatter}

\section{Introduction}

\begin{figure}[!h]
  \centering
  \raisebox{0cm}{(a)}\includegraphics[width=.9\linewidth,angle=180]{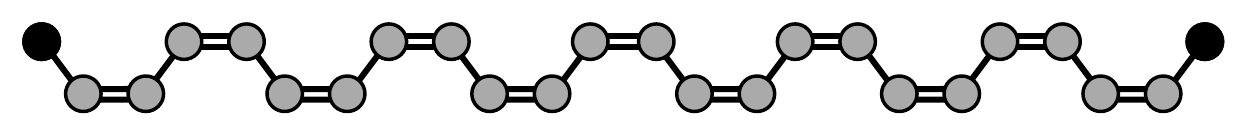}\\[-2mm]
  \raisebox{0cm}{(b)}\includegraphics[width=.9\linewidth,angle=180]{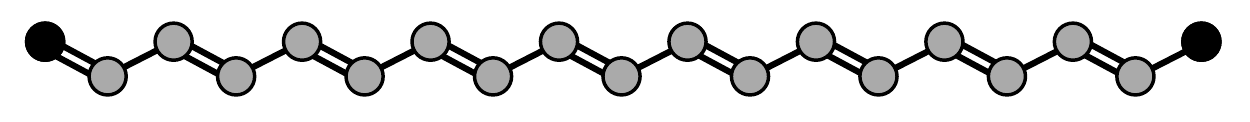}\\[1mm]
  \raisebox{1.1cm}{(c)}\includegraphics[width=.9\linewidth]{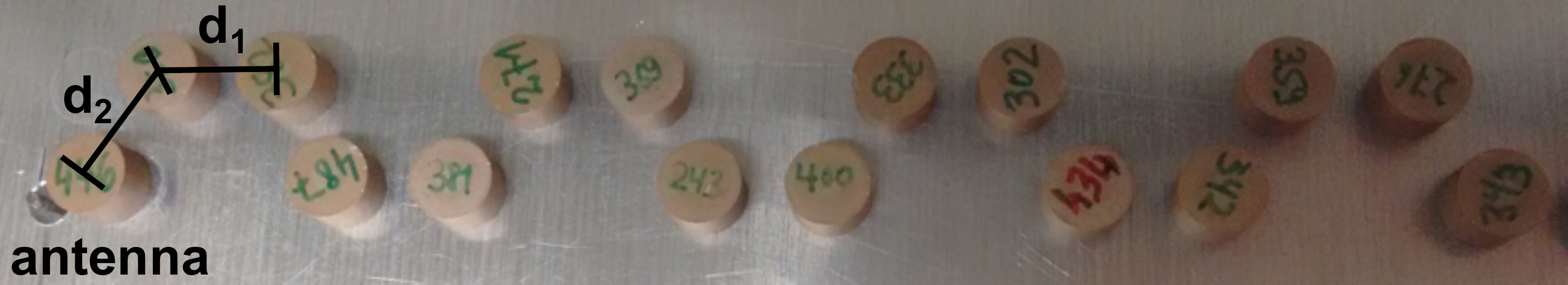}\\[1mm]
  \raisebox{1.1cm}{(d)}\includegraphics[width=.9\linewidth]{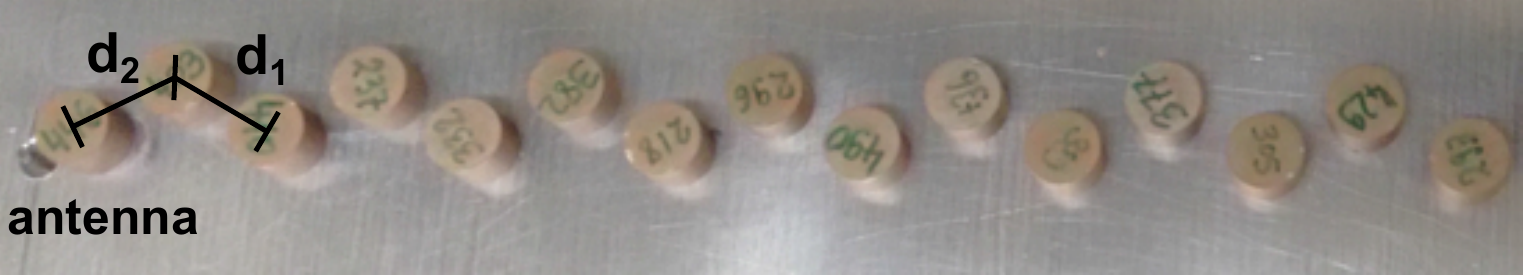}
  \caption{Structure of polyacetylene chains in \textit{cis}-geometry (a) with $N_d=11$ dimers and
    $N_c=2$ additional edge atoms, which are not part of the dimers, and in \textit{trans}-geometry
    (b) with $N_d=9$ dimers and a single edge atom ($N_c=1$) at the left chain end. The resonators
    correspond to carbon atoms, the hydrogen atoms are not shown. The blue shaded resonators at the
    chain ends indicate the resonators to which the contacts (or antennas) are coupled to study the
    transport. The cis-chain corresponds to the armchair shape, whereas the trans-chain corresponds
    to a zigzag shape. In (c) and (d) photos of the microwave experiment to emulate polyacetylene
    chains for the cis- and trans-geometry are shown, respectively. On the left hand side the
    antenna on the bottom plate is seen, whereas the antenna on the right hand side is mounted to
    the top plate (not shown).}
  \label{fig:1}
\end{figure}

The aim of further miniaturization of electronic devices has led in recent years to the question, if
it is possible to shrink down the individual active element to a single molecule. This question
stimulated the research field of molecular electronics, see \cite{Cuevas2010} and references therein
for an overview. One approach is based on carbon nanotubes, which for certain geometric parameters
(i.e. tube diameter or number of windings) show a band gap and hence, can be used as transistors
\cite{Tans1998, Javey2003, Martel1998, Charlier2007}. A recent milestone has been the realization of
a carbon nanotube computer \cite{Shulaker2013}. However, a drawback of this approach is that after
the growth of the nanotubes the metallic carbon nanotubes have to be separated from the
semi-conducting ones. An alternative approach could be to use individual polyacetylene chains, see a
sketch in \fig{1} (a,b), which are predicted to have a band gap \cite{Su1979, Heeger1988,
  Heeger2001, Schnurpfeil2007, Nozaki2008, Nozaki2010}. However, most of the experimental work has
been done with thin films of polyacetylene chains \cite{Shirakawa1995, Shirakawa2001}. Transport
experiments with \textit{individual} polyacetylene chains have, to the best of our knowledge, not
been performed yet. However, new developments in microwave experiments allow to emulate
experimentally a tight-binding model, which has proven successful in studies of graphene
\cite{Bellec2013, Bellec2013b}. These microwave experiments, which are performed here for the first
time on molecular structures, can measure the transport of microwaves through one or two dimensional
tight-binding systems. The microwave transmission corresponds to the ballistic single electron
transport in mesoscopic physics or in molecules. Electron-electron interaction, which is present to
some extent in real molecules, cannot be emulated by the microwave experiment.

In this paper we explore the transport in polyacetylene like systems. In long chains a transport gap
is observed, which is expected due to the dimerization of the chain, i.e. the chain is composed of
unit cells of two carbon atoms (dimers). Additional edge atoms add edge localized resonance states
within the gaps, which become important for short chains but are negligible for long chains.

\section{System: Polyacetylene chains}

The setup, which we use to emulate both the \textit{cis}- (armchair) and \textit{trans}- (zigzag)
isomers of polyacetylenes of various lengths, is shown in \fig{1} (c,d). The studied chains consist
of $N_d$ dimers, which we here define via the double bonds, with $N_c$ additional edge atoms at the
chain ends, which do not belong to dimers. The chains have altogether $N=2N_d+N_c$ atoms. For
example, a cis-chain consisting of $N_d=11$ dimers with $N_c=2$ additional edge atoms, one at each
end of the chain, is shown in fig{1}(a). A trans-chain of $N_d= 9$ dimers with a single edge atom on
the left can be seen in \fig{1} (b).

\begin{figure}
  \centering
  \includegraphics[scale=0.2]{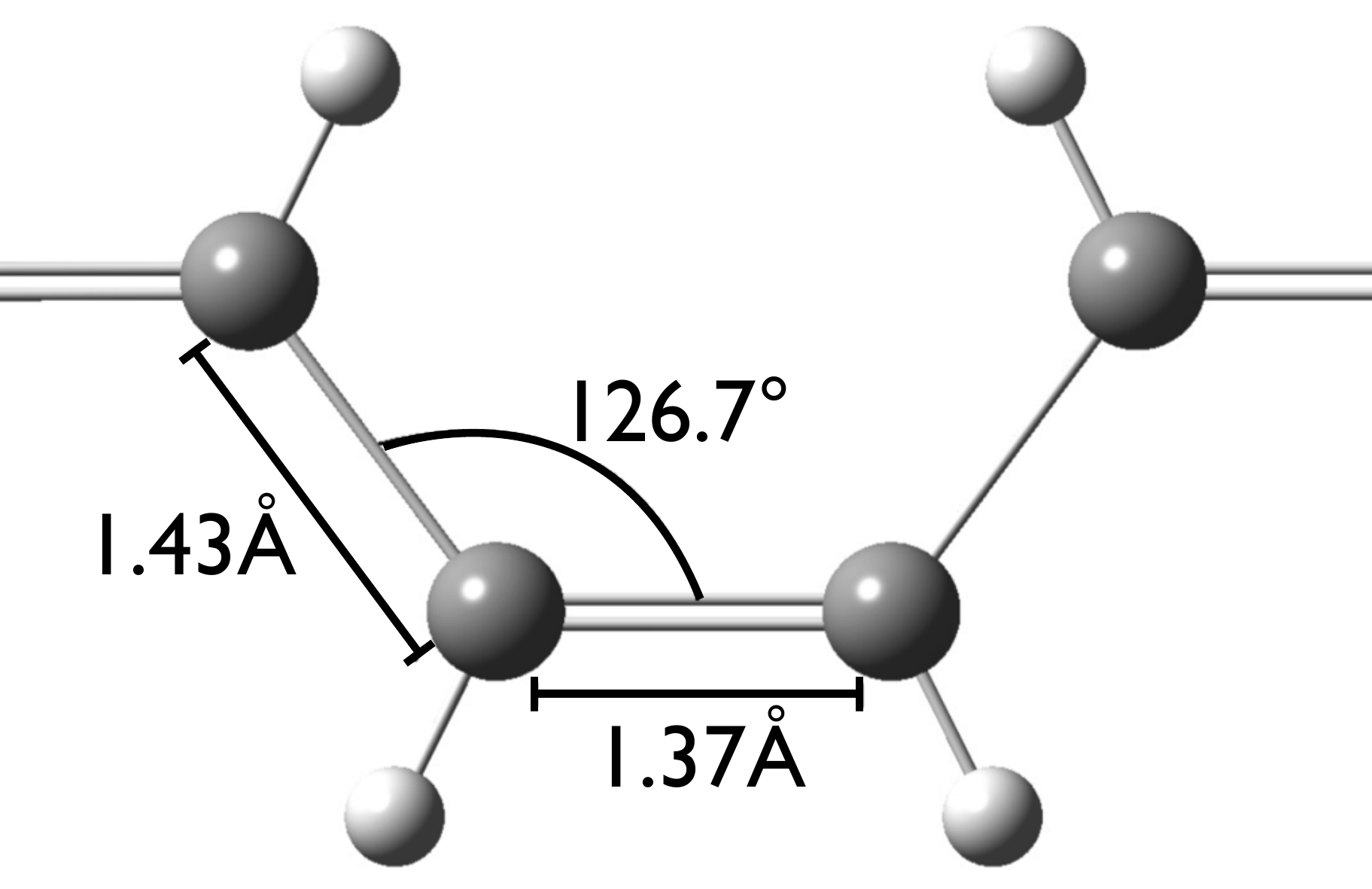}
  \hspace{3mm}
  \includegraphics[scale=0.2]{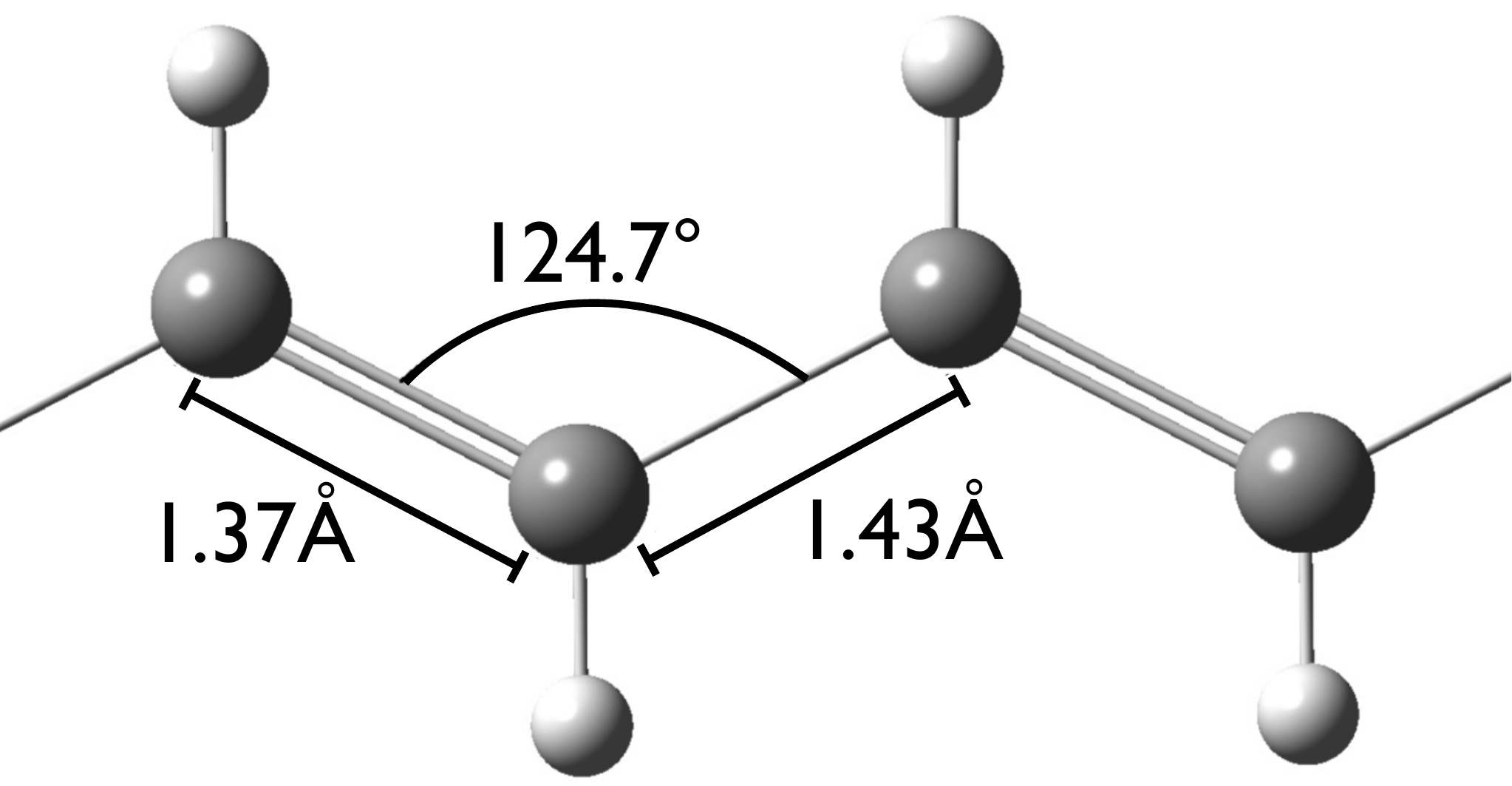}
  \caption{Stable structure of the polyacetylene molecule, optimized by the DFT method. The left
    part shows the cis-configuration (armchair) and the right part the trans-configuration (zigzag)
    with the corresponding angles and distances between the carbon atoms. The larger gray spheres
    correspond to the carbon atoms and the smaller white spheres to the hydrogen atoms, which
    saturate the dangling bonds of the carbon atoms.}
  \label{fig:2}
\end{figure}

\subsection{DFT studies}

At first, we have calculated the optimal structure of the molecule by means of density functional
theory (DFT). The key idea of DFT is that the ground-state of an interacting many-body system is a
unique functional of the electron density \cite{Hohenberg1964, Kohn1965}. The ground-state is found
by minimizing its energy with respect to the electron density. In order to perform the minimization
of the ground-state energy functional, an atomic basis and an exchange-correlation potential have to
be selected. The atomic basis is used to construct attempts to the ground-state, while the
exchange-correlation potential is taking into account the Coulomb interaction between the
electrons. It is known exactly only for the free electron gas, but accurate approximations have been
developed. This reduces the interacting many-body problem to a system of non-interacting particles
in an effective potential. Here, we use the 6-311g(d) basis set \cite{Ditchfield1971}, which
consists of Gaussian functions, and the B3LYPD hybrid exchange-correlation functional
\cite{Becke1988}. The calculations are performed by the DFT program GAUSSIAN09 \cite{Frisch2009}. A
detailed introduction into DFT can be found for example in \cite{Parr1989, Sholl2009}.

Polyacetylenes are chains of carbon atoms. The dangling bonds of the carbon atoms are saturated by
hydrogen atoms which are taken into account in our DFT studies, see \fig{2}. However, as their
contribution to the conductance can be neglected, the hydrogen atoms will not be considered in the
microwave experiment nor in the tight-binding model, see \fig{1}. In both cases, the most stable
structure is found for dimerized chains, where the nearest-neighbor distance of the carbon atoms
alternates between two different values. For the cis-chain, the horizontal bonds have a length of
$1.37 \angstrom$ while the diagonal bonds have a length of $1.43 \angstrom$ with an angle of
$126.7^\circ$, see \fig{2} (left). For the trans-chain, the bond lengths are $1.37 \angstrom$ and
$1.43 \angstrom$ with an angle of $124.7^\circ$, see \fig{2} (right). Our findings agree with other
ab initio studies of these molecules \cite{Teramae1986, Rodriguez1995}. This dimerization, which is
also known as Peierls distortion \cite{Peierls1955}, is indicated in \fig{1} and \fig{2} by
alternating single and double bonds between the carbon atoms. Homogenous chains, where the
nearest-neighbor distance of all carbon atoms is the same, do not give the energetically optimal
structure.

\subsection{Microwave experiment}

Using the techniques, developed to investigate the band structure of graphene \cite{Bellec2013,
  Bellec2013b} and to emulate relativistic systems \cite{Franco2013, Sadurni2013}, we have performed
an analogous experiment with microwave resonators to study the transport properties of the chains. A
set of identical dielectric cylindrical resonators ($5\un{mm}$ height, $4\un{mm}$ radius, refractive
index $n\approx 6$) is placed between two metallic plates. The nearest neighbor distance of the
resonators is $d_1=12.0 \un{mm}$ for the long bonds and $d_2=11.5 \un{mm}$ for the short bonds,
giving the same distance ratio $d_1/d_2$ as in the DFT calculations. Close to the chain ends, see
the blue shaded resonators in \fig{1}, antennas are placed through which microwaves (transverse
electrical (TE-)modes) can be injected and detected. The individual resonators have an isolated
resonance at $\nu_0'=6.65 \pm 0.005 \un{GHz}$. We restrict our investigation to frequencies around
$\nu_0'$, where each resonator contributes only one resonance. From now on we will use the
normalized frequency $\nu=\nu'-\nu_0'$, where the resonance frequency of the resonators is at
$0 \un{MHz}$. Photos of the experimental setup (without the metallic plate on top) are shown in
\fig{1} (c,d), and a detailed description can be found in \cite{Bellec2013b}.

\subsection{Tight-binding model}

Theoretically, we model the polyacetylene chains by the tight-binding Hamiltonian
\begin{equation}
  \label{eq:1}
  H= \sum_{\Abs{i-j}\leq d} t_{\Abs{i-j}}\ket{i}\bra{j},
\end{equation}
which provides a qualitative good description of the electronic transport in polyacetylene chains
\cite{Su1979, Heeger1988, Nozaki2010, Barford2005, Geoghegan2013}. A more complete understanding can
certainly be gained by taking into account electron correlations \cite{Kivelson1981, Baeriswyl1985},
yet the tight-binding model is often a good starting point for studies of molecular electronics
\cite{Cuevas2010}.

\begin{table}
  \centering
  \begin{tabular}{c|c|c|c|c}\hline
    cis-chain (armchair) & $t_1$ & $t_1'$ & $t_2$ & $t_3$\\\hline
    \begin{minipage}{40mm}
      \centering
      \includegraphics[scale=0.24]{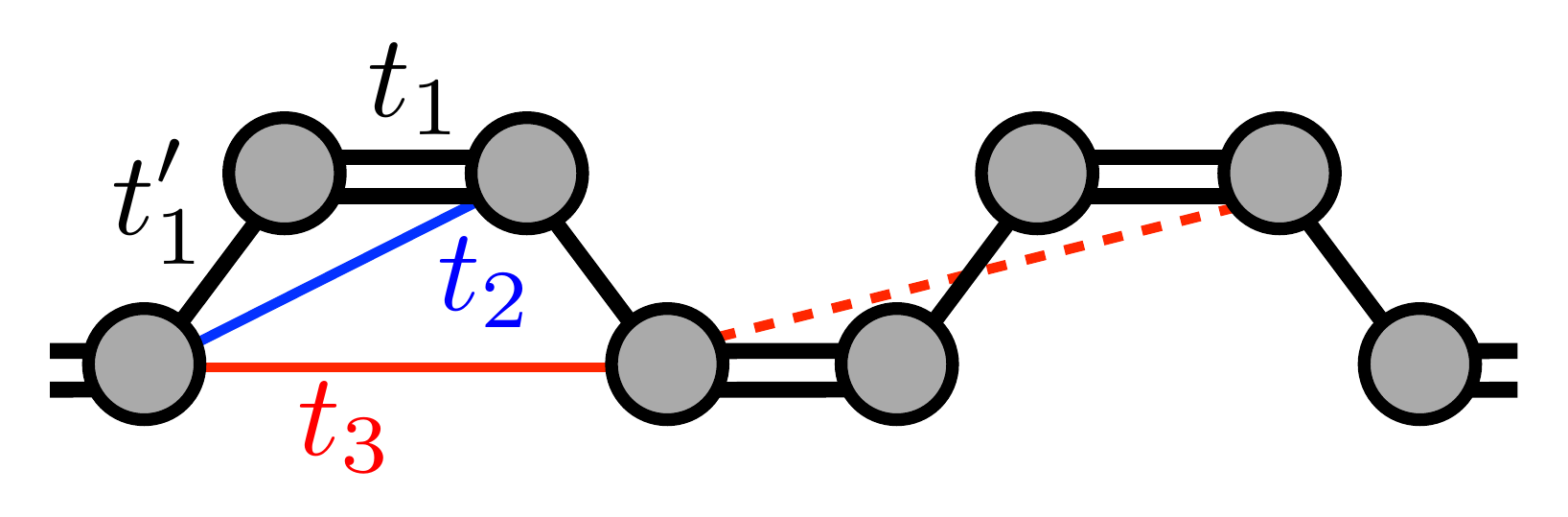}
    \end{minipage} & 43.9 & 36.2 & 3.2 & 3.1\\\hline
    trans-chain (zigzag) & & & & \\
    \hline
    \begin{minipage}{40mm}
      \centering
      \includegraphics[scale=0.23]{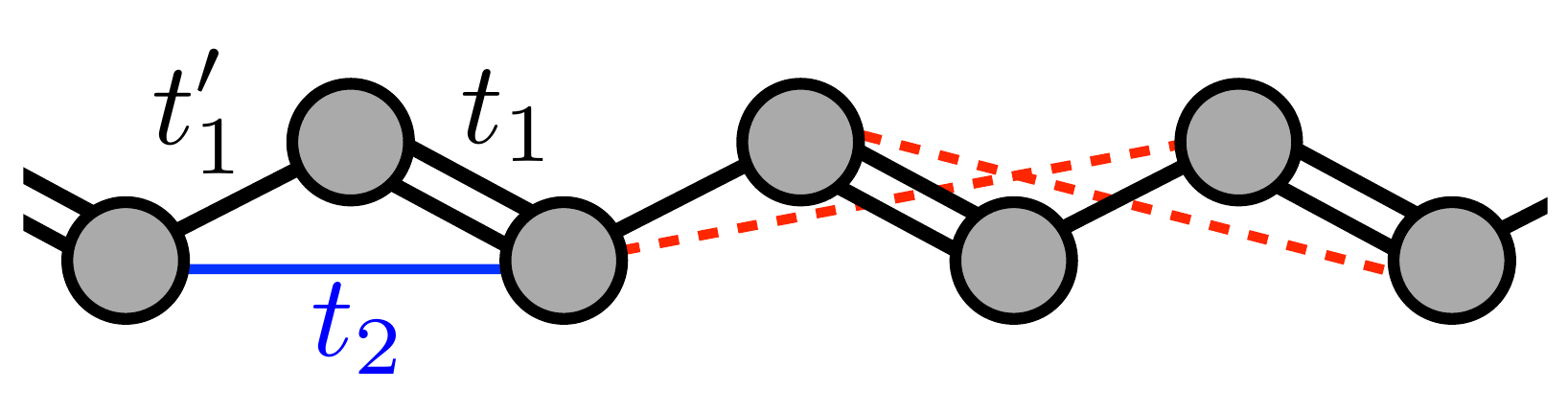}
    \end{minipage}& 43.9 & 36.2 & 4.2 & ---
  \end{tabular}
  \caption{Coupling strength (in MHz), which are used to fit our calculation to the experiments. The
    couplings between first nearest neighbors are identical for both types of chains. The two isomers
    differ in the coupling strengths between second and third nearest neighbors. Interactions to higher
    nearest neighbors, see dashed lines in the figures, can be ignored due to their large distance and
    the screening of surrounding atoms. Notice that $t_3$ is not existing in the trans-chain
    (zigzag) configuration.}
  \label{tab:1}
\end{table}

\begin{figure}
  \centering
  \includegraphics[scale=0.65]{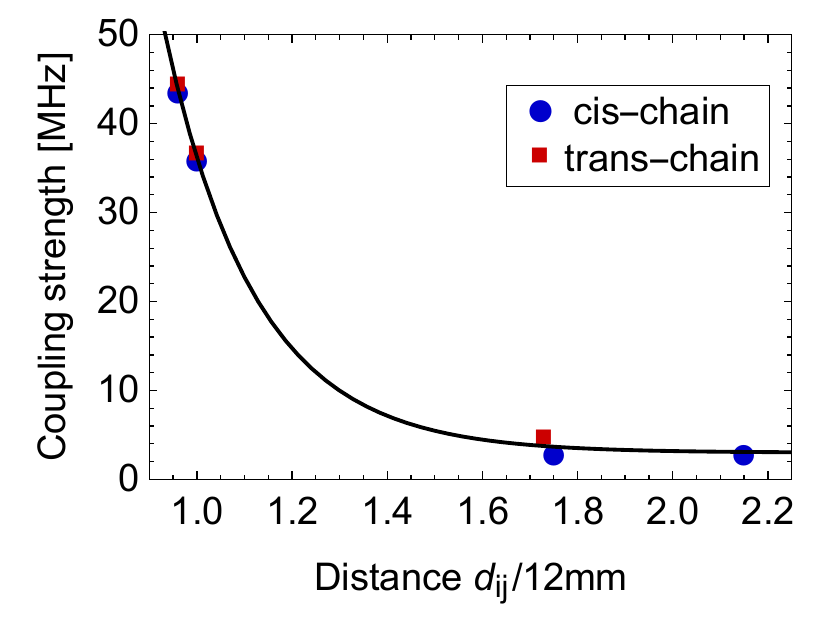}
  \caption{The coupling strength between between the resonators decays approximately exponentially
    with the distance $d_{ij}$ of the resonators. The solid line is given by
    $t=6031\un{MHz}\exp(-5.2 d_{ij}/12\un{mm})+2.99\un{MHz}$.}
  \label{fig:3}
\end{figure}

The location of the isolated spectral peaks for the smallest cis- and trans-chains in our
experiments are used to fit the onsite energies and the coupling parameters, i.e. the part of the
spectrum for the smallest molecules which is uniquely determined by our measurement is used to fix
the non-zero couplings for each of the two chains. This is in keeping with the standard technique of
extracting a Hamiltonian from a ``polyad'' for molecules \cite{Troellsch2001, Jung2002}. Note, that
we actually have polyads here because similar other spectra will appear for different frequency
ranges, yet they are well separated from the spectra we study. The thus obtained coupling strengths
$t_i$ are listed in \tab{1} and depicted in \fig{3}. The sum in \eq{1} takes into account
interactions between the resonators up to the 3rd nearest neighbors (definition see \tab{1}). Have
in mind that we define the $n$'th nearest neighbor by its index difference $n=\Abs{i-j}$ along the
chain and not by the real distances $d_{ij}=\Abs{\vec{r}_i-\vec{r}_j}$. As the distances $d_1$ and
$d_2$ of first nearest neighbors (1nn) are identical in both types of chains, see \fig{2}, also
their couplings $t_1$ and $t_1'$ are the same for both isomers. As the nearest neighbor interactions
are the dominant ones the measured and calculated transmission shows \textit{qualitatively} the same
main features for cis-(armchair) or trans-(zigzag) geometry. However, subtle differences can be
noticed due to interactions between resonators, which are farther away. Due to the different bond
angles in the two isomers, see \fig{2}, the couplings between second nearest neighbors (2nn, $t_2$)
are different. Polyacetylene chains have two types of third nearest neighbors. Third nearest
neighbors of the first type can be neglected due to their large distance and the strong screening by
the closer resonators, see the dashed lines in the figures of \tab{1}. Third nearest neighbors of
the second type ($t_3$) exist only in cis-chains (armchair) and cannot be neglected to understand
the experimental data. Higher nearest neighbors couplings have been ignored in both chains as
well. In \sec{res1}, we discuss effects on the transmission which would be observed if, for example,
the third nearest neighbors in the cis-chains would not be taken into account. The coupling strength
decays approximately exponentially with the distance of the resonators \cite{Bellec2013b}, see
\fig{3}. In first approximation, an exponential decay is also expected in real polyacetylene chains,
where the coupling strength is determined by the overlap of the atomic wave functions.

\subsection{NEGF method}

In order to calculate the transmission through the chains, we apply the non-equilibrium Green's
function (NEGF) method \cite{Datta1997, Cuevas2010}. The Green's function of the chain is defined as
\begin{equation}
  \label{eq:3}
  G= \Big[\nu -H -\Sg_1 -\Sg_N -\Sg_{\text{abs}} -\Sg_{\text{dis}} \Big]^{-1},
\end{equation}
where $\nu$ is the microwave frequency or, in analogy, the electron energy. The influence of the
antennas (or source and drain contacts) coupled to the chain ends (blue shaded atoms number $1$ and
$N$ in \fig{1}) is taken into account by an imaginary self-energy
\begin{equation}
  \label{eq:4}
  \Sg_{1}= -\I \eta \ket{1}\bra{1}
  \quad \text{and} \quad
  \Sg_{N}= -\I \eta \ket{N}\bra{N},
\end{equation}
where the coupling strength $\eta= 1 \un{MHz}$ is adjusted to the experiment. The absorption, which
is present in the experiment, is modeled by the imaginary potential (or self-energy)
\begin{equation}
  \label{eq:5}
  \Sg_{\text{abs}}= -\I \gamma(N) \sum_{i=1}^N \ket{i} \bra{i}.
\end{equation}
We obtain best agreement between the experiments and the calculations assuming a linear decay of the
absorption
\begin{equation}
  \label{eq:10}
  \frac{\gamma(N)}{\text{MHz}}= 1.17 -0.013N.
\end{equation}
In the experiment, some degree of disorder cannot be avoided completely due to the uncertainty of
the resonance frequency of the resonators and the uncertainty of their positions. In the
calculations, disorder is taken into account by a random potential (or self-energy)
\begin{equation}
  \label{eq:11}
  \Sg_{\text{dis}}= \sum_{i=1}^N \eps_i \ket{i} \bra{i},
\end{equation}
where the $\eps_i$ are chosen from a Gaussian distribution which is cut at its full width half
maximum, which corresponds approximately to the experimentally observed distribution and the used
selection rule. We consider the standard deviation $\sg=5 \un{MHz} $ and an ensemble of $10^3$
realizations.

The transmission between source and drain is then given by
\begin{equation}
  \label{eq:6}
  T(\nu)= 4\Tr{\Im{\Sg_1}G\,\Im{\Sg_N}G^\dagger}.
\end{equation}
Note that in absence of interactions, as in this study, the NEGF method is equivalent to a
scattering or transfer matrix method \cite{Datta1997}.

\section{Results}

Our experimental and theoretical results are shown in \fig{4} for the cis-polyacetylene chains and
in \fig{5} for the trans-polyacetylene chains. The Green's function calculations (red dashed curves)
agree qualitatively with the experimental data (blue solid curves), in particular for short
chains. The differences between the resonance frequencies are within the width of the resonances and
apart from a few resonances their heights vary on the 50\% level.

\subsection{Transport band gap due to dimerization}

\begin{figure}[t]
  \centering
  {\small  cis-chains (armchair)}\\[-1mm]
  \mbox{\includegraphics[scale=0.52]{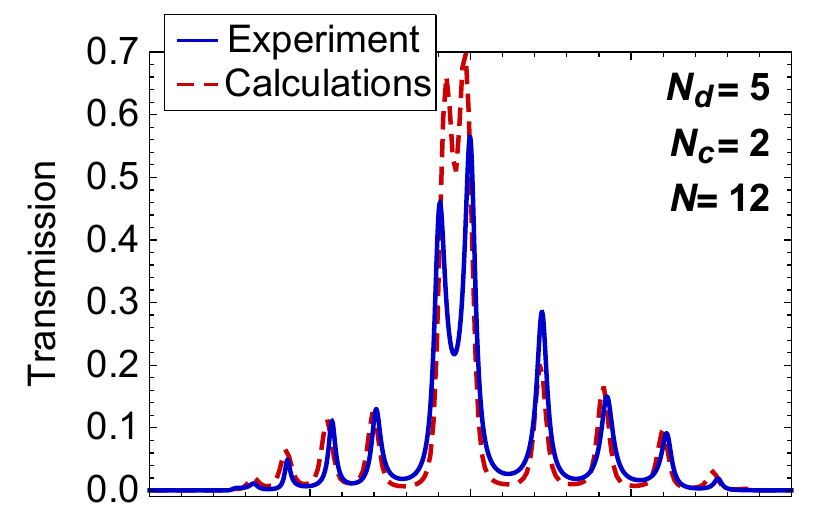} \hspace*{-1mm} \includegraphics[scale=0.52]{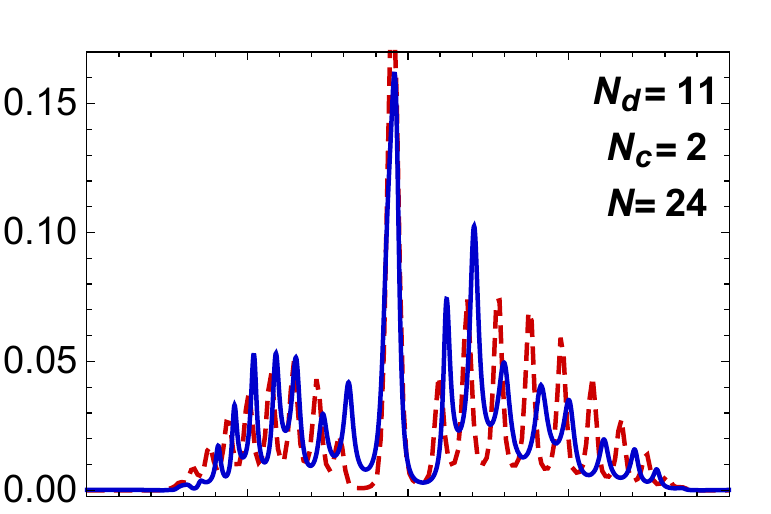}}\\[2mm]
  \mbox{\includegraphics[scale=0.52]{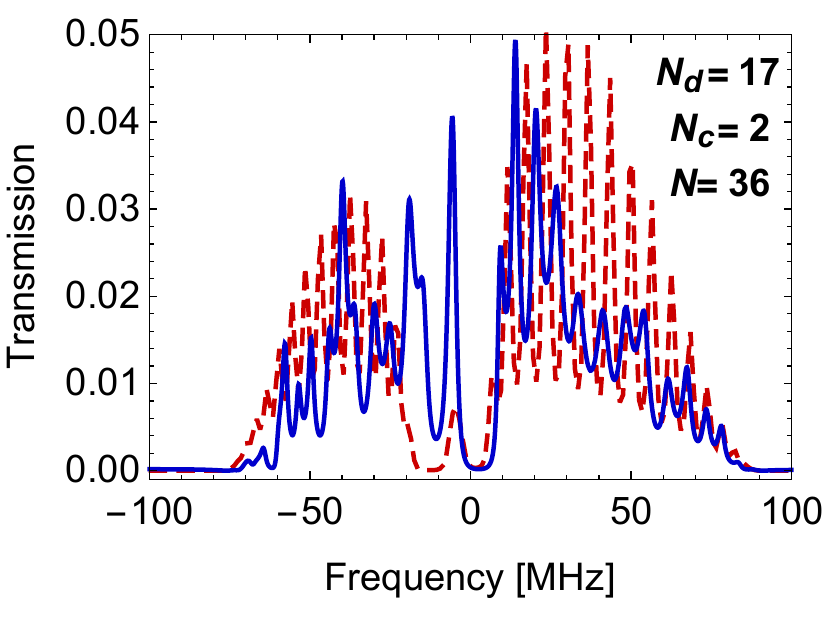} \hspace*{-1mm} \includegraphics[scale=0.52]{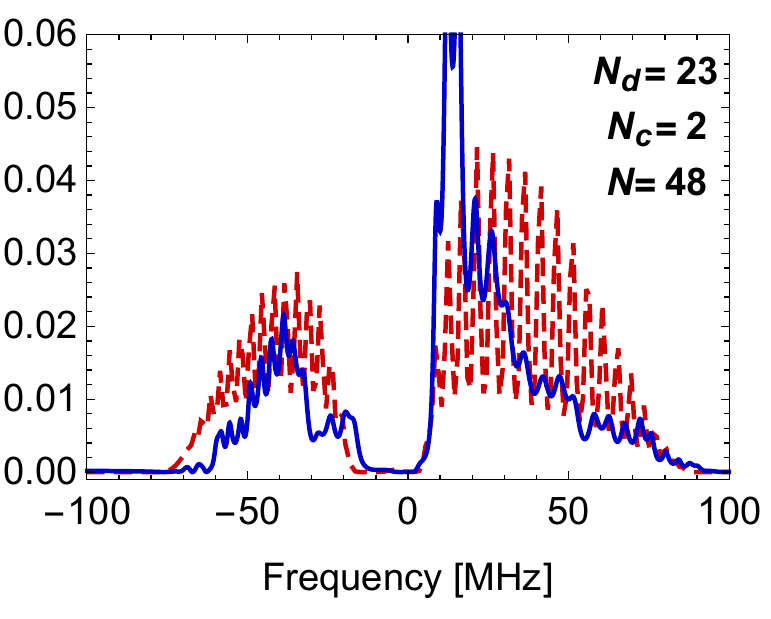}}\\[-4mm]
  \caption{Transmission as a function of the microwave frequency through cis-polyacetylene chains
    (armchair) of increasing length $N$. The blue-solid curves give the experimental data while the
    red-dashed curves represent the Green's function calculations. The short chains show distinct
    peaks, in particular around the frequency $\nu=0$. With increasing chain length a band gap
    around this frequency opens.}
  \label{fig:4}
\end{figure}

\begin{figure}[t]
  \centering
  {\small  trans-chains (zigzag)}\\[-1mm]
  \mbox{\includegraphics[scale=0.52]{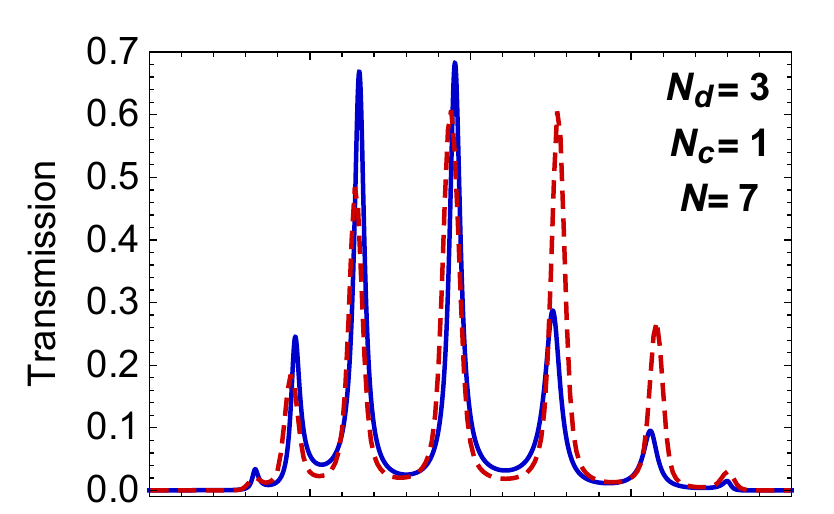} \hspace*{-1mm} \includegraphics[scale=0.52]{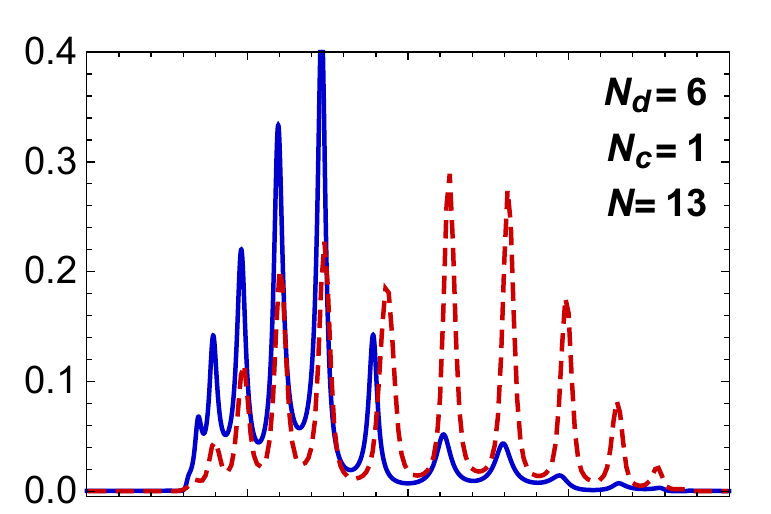}}\\[2mm]
  \mbox{\includegraphics[scale=0.52]{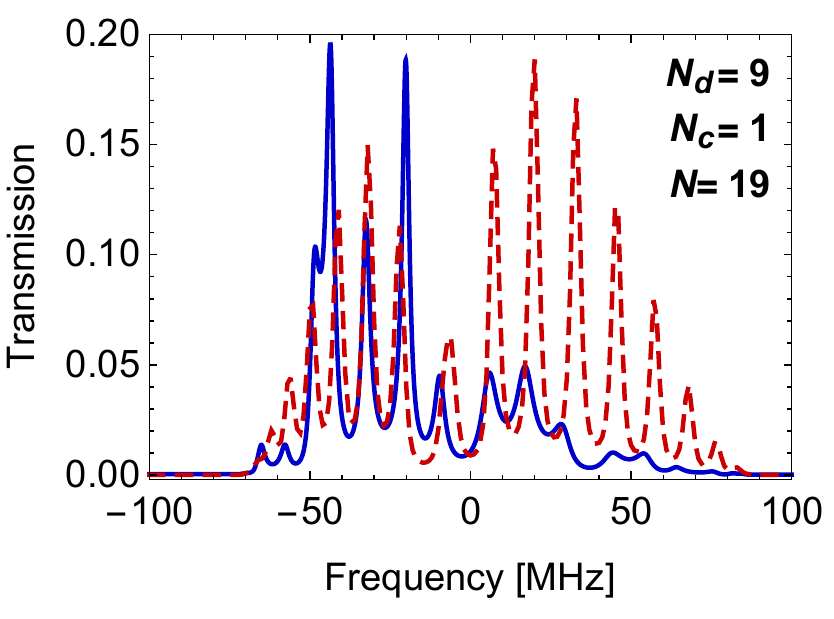} \hspace*{-1mm} \includegraphics[scale=0.52]{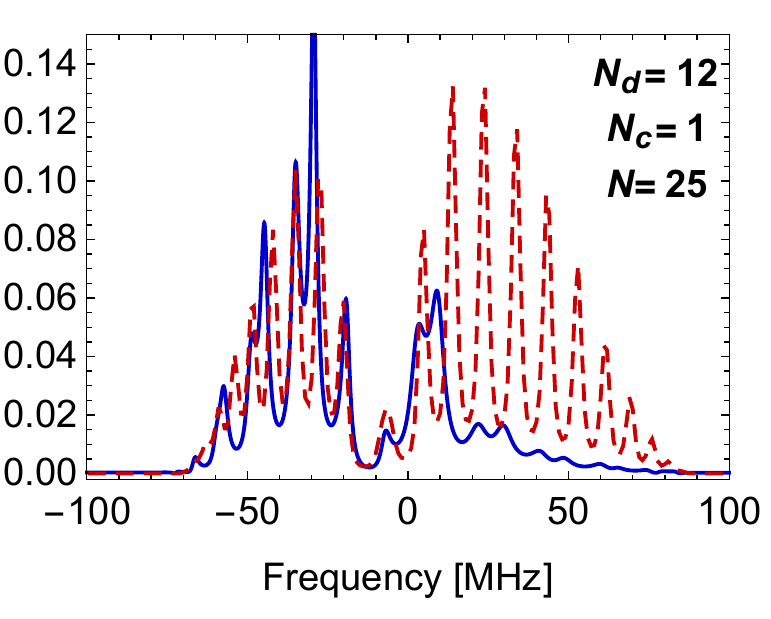}}\\[-4mm]
  \caption{Transmission through trans-polyacetylene chains (zigzag). Experimental data are shown by
    blue-solid curves, calculations by red-dashed curves. Distinct transmission resonances can be
    observed. The opening of a band-gap with increasing chain length can be seen only slightly
    because the total length of the trans-chains is less compared to the cis-chains studied in
    \fig{4}.}
  \label{fig:5}
\end{figure}

The short chains still show individual resonance peaks, which are located approximately at the
eigenenergies of the closed tight-binding Hamiltonian \eq{1}. When the chain length is increased,
these resonances are less pronounced and two conduction bands arrise, which are separated by a band
gap around the frequency $\nu=0$. The band gap exists in the cis- as well as in the trans-chains but
it is more pronounced in the former configuration. This can simply be attributed to the fact that
the number of atoms and hence the chain length is much larger in the cis- than in the
trans-chains. Numerical calculations confirm that the band gap becomes more pronounced also in the
trans-configuration when the chain length is increased.

\begin{figure}
  \centering
  {\small \hspace{5mm} cis-chains (armchair) \hspace{15mm} trans-chains (zigzag)}\\[-1mm]
  \mbox{\includegraphics[scale=0.54]{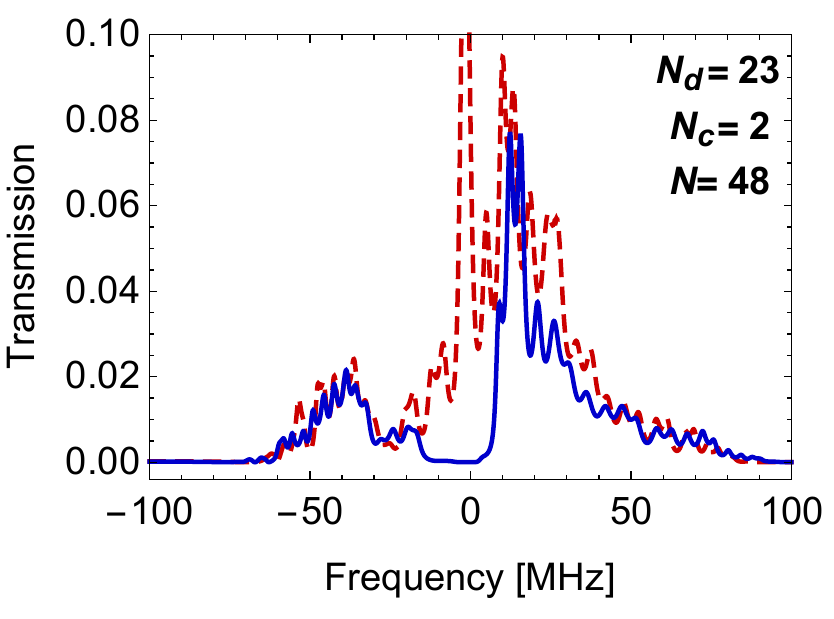} \hspace*{-1mm} \includegraphics[scale=0.54]{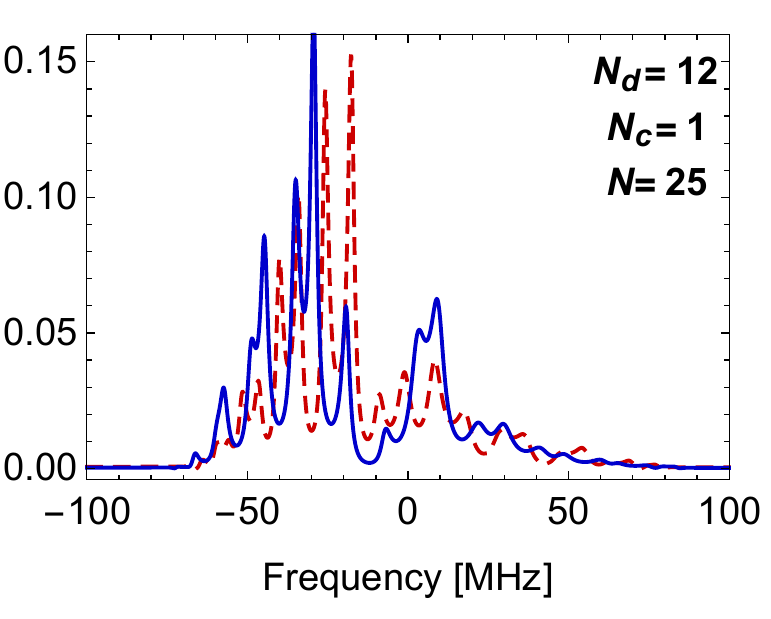}}\\[-4mm]
  \caption{Experimentally measured transmission through dimerized chains (blue solid curves) and
    homogenous chains (red dashed curves). The experiment shows that the dimerized chains have a
    band gap, the homogeneous chains not.}
  \label{fig:6}
\end{figure}

Additionally, \fig{6} shows the experimentally measured transmission through dimerized chains
(blue-solid curves) and homogenous chains (red-dashed curves), where the bond lengths are all
equal. In this case the chains are composed by units of only one carbon atom (monomer). The
homogenous chain does not show a band-gap but a conduction band ranging from $-70 \un{MHz}$ to
$90\un{MHz}$ and hence, confirms that the opening of the band gap is due to the dimerization of the
chain. The opening of the band gap due to the dimerization can be understood by considering an
infinitely long chain with first nearest neighbor interaction only. Such a first nearest neighbor
tight-binding model is known in chemistry also as the H\"uckel model \cite{Huckel1931, Murrell1985}.
The tight-binding Hamiltonian of this chain can be written as
\begin{equation}
  \label{eq:7}
  H_{\infty}= \sum_{i=-\infty}^\infty t\ket{i,1}\bra{i,2} +t'\ket{i,2}\bra{i+1,1} +\text{H.c.},
\end{equation}
where $t$ and $t'$ are the alternating coupling strengths in the dimerized chain. The quantum states
of the dimers are denoted by $\ket{i,1}$ and $\ket{i,2}$. Applying a Fourier transform, we obtain a
simpler $2 \times 2$ Schr\"odinger equation for the dimer state in momentum space
\begin{equation}
  \label{eq:8}
  \begin{pmatrix}
    0 & t +t' e^{\I k}\\
    t +t' e^{-\I k} & 0
  \end{pmatrix}
  \begin{pmatrix}
    k,1\\
    k,2
  \end{pmatrix}
  =
  \eps(k)
  \begin{pmatrix}
    k,1\\
    k,2
  \end{pmatrix}.
\end{equation}
Its eigenvalues are
\begin{equation}
  \label{eq:9}
  \eps(k) = \pm \sqrt{\Delta^2+ 4tt'\cos^2(k/2)},
\end{equation}
where $2\Delta= 2\Abs{t-t'}$ gives the band gap. For the chains above, we have
$2 \Delta= 2\Abs{43.9-36.2} \un{MHz} \approx 15.4 \un{MHz} $, which agrees with the gap of around
$15\un{MHz}$ seen in \fig{4} and \fig{6} for the experimental data. Have in mind that in the
calculation the natural broadening is not taken into account, which leads to an reduction of the
observed band gap. Note that the above equations are also valid for the homogenous chain ($t=t'$),
where the band gap $2 \Delta=0$ disappears.

\subsection{Importance of higher-neighbor couplings}
\label{sec:res1}

\begin{figure}
  \centering
  \mbox{\includegraphics[scale=0.54]{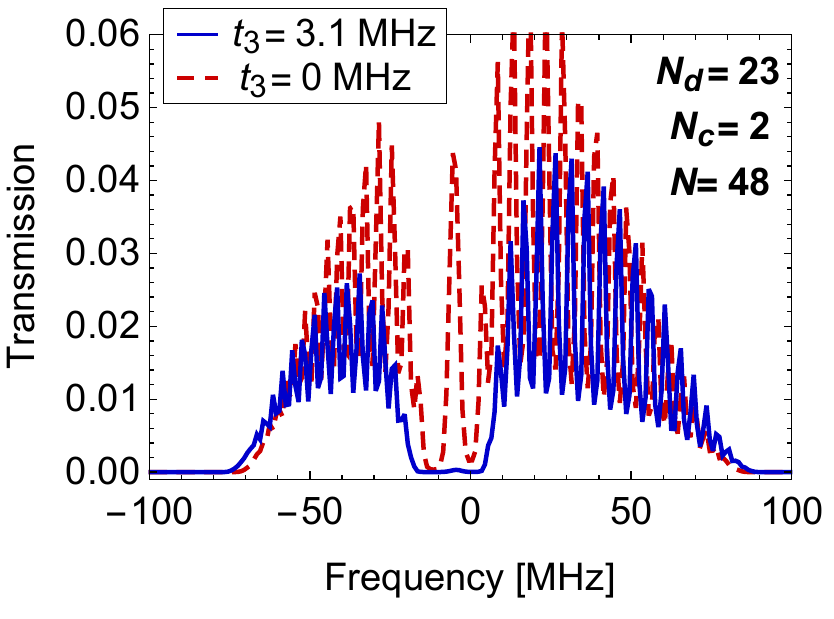} \hspace*{-1mm} \includegraphics[scale=0.54]{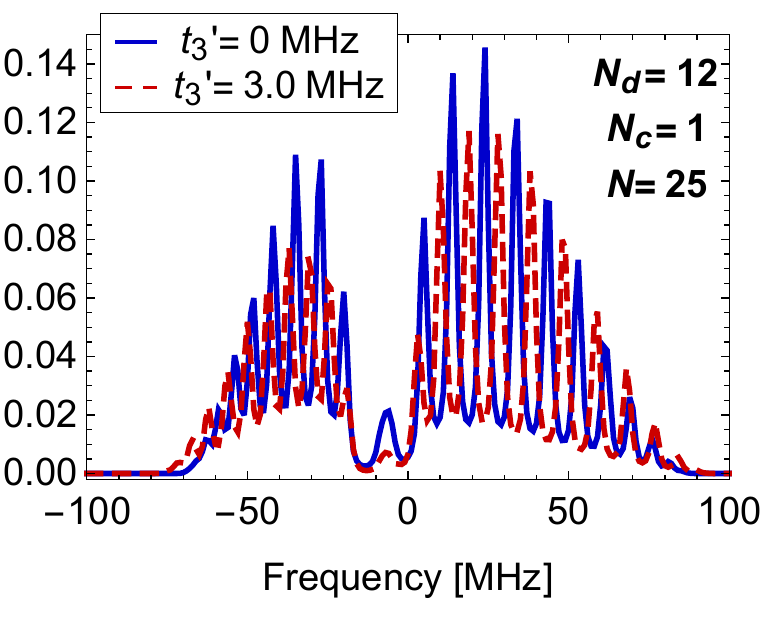}}\\[-4mm]
  \caption{Left: Calculated transmission through the longest studied cis-polyacetylene chain. If
    only couplings up to second nearest neighbors would be taken into account (red-dashed curve), a
    distinct resonance peak in the center of the band gap would arrise, which is not present if also
    third nearest neighbors are considered (blue-solid curve) nor in the experimental data (see
    \fig{4}). Right: For the trans-chain contrary behavior is observed. If interactions to third
    nearest neighbors are taken into account by means of \fig{3}, the small resonance peak at
    $\nu=-5 \un{MHz}$ is suppressed which does not agree with the experimental data in \fig{5}.}
  \label{fig:7}
\end{figure}

We would like to emphasize that the inclusion of the third neighbor coupling $t_3$ is crucial to
obtain the gap for the cis-chain as can be seen in \fig{7}. If the $t_3$ is set to zero in the
numerical calculation, a distinct transmission peak occurs at $\nu= -5 \un{MHz}$, which is neither
present in the experiment (see \fig{4} (lower right)) nor in the calculation including $t_3$
appropriately. On the other hand side for the trans-chain the contrary behavior is observed once we
include couplings to third nearest neighbors extracted from \fig{3}. The small resonance peak close
to $\nu=-5\un{MHz}$ is suppressed not agreeing with the experimental observation in \fig{5} (lower
right). We attribute this to the fact that the coupling is suppressed by the shielding of the other
resonators (see \tab{1}).

\subsection{Effect of edge atoms}

\begin{figure}
  \centering
  \mbox{\includegraphics[scale=0.54]{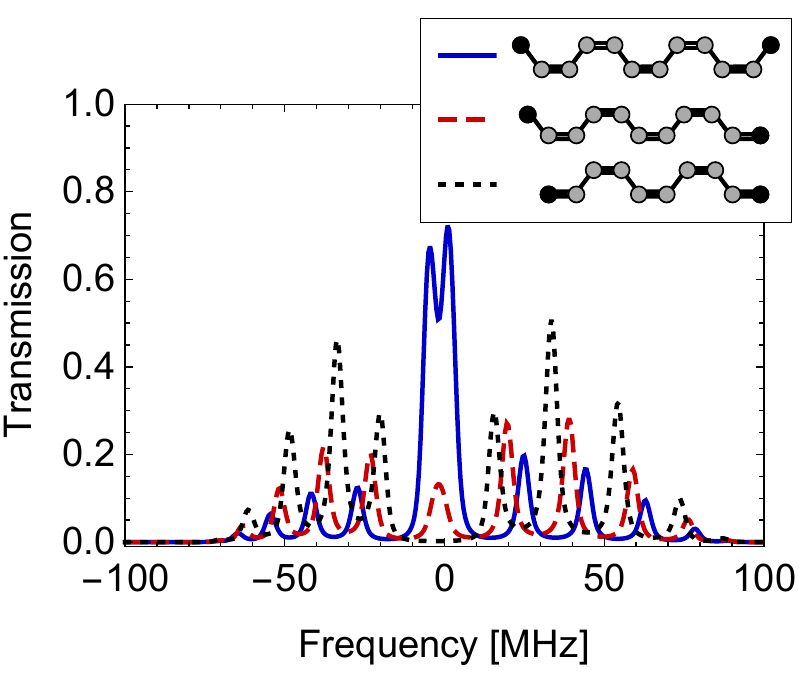}  \hspace*{-1mm} \includegraphics[scale=0.54]{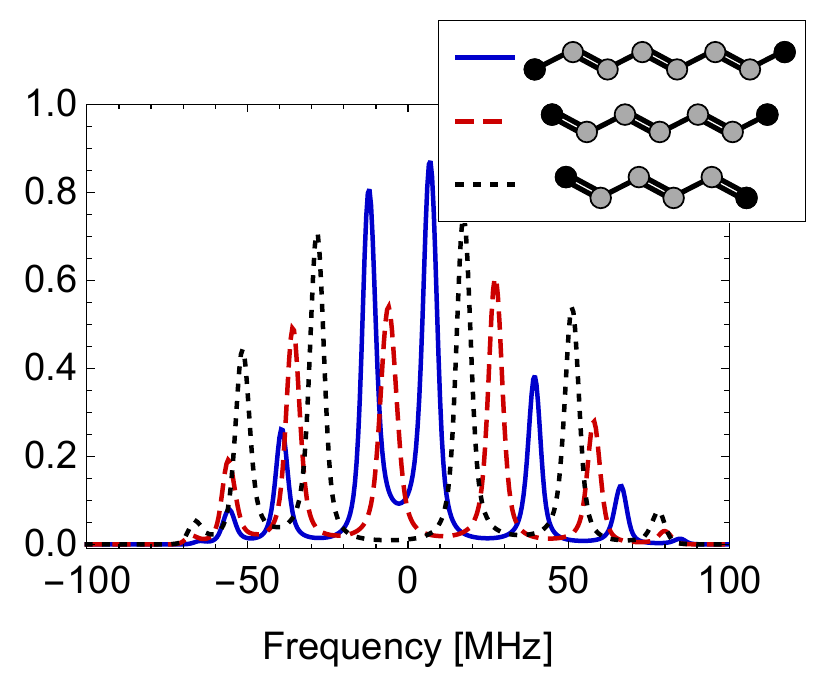}}
  \caption{Calculated transmission through the short chains sketched in the inset. In a short chain
    of dimers (black-dotted curves) a band gap or at least a broad transmission minimum can be
    observed. When additional edge atoms are attached to the chain ends (red-dashed and blue-solid
    curves), resonance peaks arrise in the band gap around $\nu=0$. These peaks are due to two edge
    state, which are localized symmetrically and anti-symmetrically at both chain ends.}
  \label{fig:8}
\end{figure}

\Fig{8} shows the calculated transmission through the short chains sketched in the inset. When the
chain consists only of dimers without any additional edge atom (i.e. $N_c=0$) also in short chains a
band gap or at least a broad transmission minimum can be observed (black-dotted curves). When an
additional edge atom is attached to one of the chain ends $N_c=1$, a distinct resonance peak is
induced in the center of the band gap around $\nu=0$ (red-dashed curves). This resonance peak
belongs to an exponentially decreasing edge state which is localized at the chain end. Due to its
localization this edge state is non-conductive in long chains and the corresponding resonance peak
disappears. When edge atoms are attached to both chain ends, two peaks are observed in the
transmission (blue-solid curves). These peaks belong to two states which are localized symmetrically
or anti-symmetrically at both ends.  With increasing chain length the splitting of these peaks and
their height decrease, until they finally do not contribute to the transmission any more. The
additional resonance peaks due to edge atoms can be seen clearly in the first row of \fig{4} and
\fig{5}. Our numerical calculations show that the same effects appear also in pure dimer chains
(shown at the bottom of the inset of \fig{8}) if the contacts (or antennas) are not coupled to the
blue shaded atoms but moved one atom inside the chain. Note that the observed resonance peaks are
not an even-odd parity effect of the total chain length, which can be observed also in molecular
chains \cite{Kim2002}.

\section{Conclusions}

Summarizing we have presented microwave experiments to emulate the transport in individual
polyacetylene chains, where the optimal structure of the molecule has been obtained by the DFT
method. The experiments have been accompanied with tight-binding transport calculations using the
NEGF method. Both, our experiments and calculations confirm that dimerization produces a band gap in
these chains. If the chains are sufficiently long, a band gap appears for cis- and trans-chains for
all possible configurations of edge atoms and contact positions. This has been seen experimentally
for a single non-dimerized edge and has been confirmed numerically for other edge and contact
configurations. Thus, returning to the motivation given at the beginning of this article, long
polyacetylene chains might indeed replace carbon nanotubes in molecular transistors. Moreover, we
have also shown that short dimer chains have a pronounced transmission minimum, if no additional
edge atoms are attached and the contacts are at the chain ends. Thus, at the expense of having to
know precisely the geometry of the molecule and the way how it is coupled to the contacts, it may
well be possible to use rather short chains for novel molecular transistors. While this is probably
the first microwave emulation of a molecule that is not largely made up of benzene rings, the actual
scope for such microwave emulations is very wide. A first option would be to study local defects,
for example, a missing double bond (e.g. due to a substitution of a hydrogen atom) \cite{Su1979,
  Heeger1988, Nozaki2010}. Such defects produce resonances within the band gap, similar to the
observed resonances due to edge atoms. A single defect has been used recently to establish a robust
topological mode, which could be used for quantum computation \cite{Schomerus2013,
  Poli2015}. Studies of other conjugated carbon systems using similar experimental methods are in
process. The successful emulation of boron--nitrogen sheets \cite{Barkhofen2013} suggest that a
larger variety of molecules can be studied with similar techniques.

\section*{Acknowledgments}
Financial support from CONACyT research grant 219993 and PAPIIT-DGAPA-UNAM research grants IG100616
and IN114014 is acknowledged. T.S. acknowledges a postdoctoral fellowship from DGAPA-UNAM. J.A.F.-V
acknowledges financial support from CONACyT project CB2012-180585. T.H.S. and J.A.F.-V. are grateful
for the hospitality regularly received at the LPMC. We acknowledge extensive use of the MIZTLI super
computing facility of DGTIC-UNAM under project SC15-1-S-20.

\section*{References}

\bibliographystyle{elsarticle-num}
\bibliography{poac-16-06-15}

\end{document}